\documentclass[sigconf]{acmart}

\renewcommand\footnotetextcopyrightpermission[1]{} 
\usepackage{todonotes}
\usepackage{graphicx}
\usepackage{url}
\usepackage{subfig}
\usepackage{color}
\usepackage[LGR,T1]{fontenc}
\usepackage{algorithm}
\usepackage{algpseudocode}
\usepackage{pifont}
\usepackage{amstext,amsmath,amssymb,amsfonts,amsbsy}

\newcommand{\textgreek}[1]{\begingroup\fontencoding{LGR}\selectfont#1\endgroup}
\allowdisplaybreaks

\def\fscore{\mbox{$f\mbox{-}score$}}

\begin{document}


\title{Searching Heterogeneous Personal Digital Traces}

\author{Daniela Vianna}
\affiliation{%
  \institution{Dept. of Computer Science}
  \institution{Rutgers University}
  \city{New Jersey}
  \state{USA}
}
\email{dvianna@cs.rutgers.edu}

\author{Varvara Kalokyri}
\affiliation{%
  \institution{Dept. of Computer Science}
  \institution{Rutgers University}
  \city{New Jersey}
  \state{USA}
}
\email{v.kalokyri@cs.rutgers.edu}

\author{Alexander Borgida}
\affiliation{%
  \institution{Dept. of Computer Science}
  \institution{Rutgers University}
  \city{New Jersey}
  \state{USA}
}
\email{borgida@cs.rutgers.edu}

\author{Thu D. Nguyen}
\affiliation{%
  \institution{Dept. of Computer Science}
  \institution{Rutgers University}
  \city{New Jersey}
  \state{USA}
}
\email{tdnguyen@cs.rutgers.edu}

\author{Am\'elie Marian}
\affiliation{%
  \institution{Dept. of Computer Science}
  \institution{Rutgers University}
  \city{New Jersey}
  \state{USA}
}
\email{amelie@cs.rutgers.edu}
\renewcommand{\shortauthors}{D. Vianna et al.}

\begin{abstract}

Digital traces of our lives are now constantly produced by various connected devices, internet services and interactions. Our actions result in a multitude of heterogeneous data objects, or traces, kept in various locations in the cloud or on local devices. Users have very few tools to organize, understand, and search the digital traces they produce. We propose a simple but flexible data model to aggregate, organize, and find personal information within a collection of a user's personal digital traces. Our model uses as basic dimensions the six questions: what, when, where, who, why, and how. These natural questions model universal aspects of a personal data collection and serve as unifying features of each personal data object, regardless of its source. We propose indexing and search techniques to aid users in searching for their past information in their unified personal digital data sets using our model. Experiments performed over real user data from a variety of data sources such as Facebook, Dropbox, and Gmail show that our approach significantly improves search accuracy when compared with traditional search tools.  

\end{abstract}

\maketitle

\vspace{-0.05in}
\section{Introduction}
\label{sec:introduction}

Digital traces of our lives are constantly being produced and saved by users, either actively in files, emails, social media interactions, multimedia objects, calendar items, contacts, etc., or passively via various applications such as GPS tracking of mobile devices, records of usage, records of financial transactions, web search records or quantified self-sensor usage. These ``personal digital traces'' are different from traditional personal files; they are typically (but not always) smaller, heterogeneous, and accessible through a wide variety of different portals and interfaces, such as web forms, APIs or email notifications; or directly stored in files used by apps on our devices. These traces reflect a chronicle of the user's life, keeping record of where the user went, who the user interacted with (online or in real-life), what the user did, and when. However, the large quantity of personal data available, and the fact that data is stored in multiple decentralized systems, in heterogeneous formats, makes it challenging for users to interact with their data and perform even simple searches.

Our goal is to give back to individual users easy and flexible access to their own data.  In~\cite{DataExtraction} we proposed an extraction tool that implements access to a variety of data sources, retrieving the decentralized data and storing it in a single database. Personal data is highly sensitive; consequently, privacy and ethical issues have to be considered while dealing with this type of information. Due to privacy concerns, the data downloaded is stored on the user's own hard drive, and aggregate query answers that we wish to see for experimental purposes must be approved by the user. More elaborate scheme for preserving privacy in personal information management is discussed in~\cite{Abiteboul:2015}. The work discussed in this paper is developed as part of a series of tools to let user retrieve, store and organize their digital traces {\em on their own devices}~\cite{ExploreDBValia,odbase,DataExtraction}, guaranteeing some clear privacy and security benefits. 

Work in Cognitive Psychology~\cite{wagenaar86,brewer88, sevensinsmemory,PIMbook} has shown that contextual cues are strong triggers for autobiographical memories. Abowd \textit{et al.}~\cite{Abowd:1999} and Dey~\cite{Dey:2001} define context as any information that can be used to characterize {\em the situation of an entity} (person, place, object,...). This suggests that a natural way to remember and learn from past events is to include any pertinent contextual information when organizing and searching personal data. Personal information can be modeled, and indexed following six dimensions that mirror the basic interrogative words: \textit{what, who, when, where, why, and how}. Each personal digital trace is a source of knowledge. For instance, a simple Facebook post may contain enough information to identify where a user went, what they did, who they interacted with, and when. Multiple traces, from the same or different data sources, are often related to each other. The correlation between data traces can be identified through common information such as time and location. Even though multiple data traces may share common information, they may have significantly different structures. This heterogeneity presents a major challenge. Thus, in this work, we are proposing a data model that can effectively represent this heterogeneous data in a way that can aid users to find pieces of information again. 

Search of personal data is usually focused on retrieving information that users know exists in their own data set, even though most of the time they do not know in which source or device they have seen the desired information. Current search tools such as Spotlight and Gmail search are not adequate to deal with this scenario where the user has to perform the same search multiple times on different services or/and devices rather than search over just a single service. Besides, traditional searches are often inefficient as they typically identify too many matching documents. In addition to the unified data model, we are proposing scoring and searching techniques that allow personal information search over distributed data from multiple services and devices integrated in a unified data set. 

In this paper, we make the following contributions:
\begin{itemize}
\vspace{-0.15cm}
\item A unified and intuitive multidimensional data model to link and represent heterogeneous personal digital traces. The model, called \emph{w5h}, uses those six dimensions to unify features of each personal data object, regardless of its source. (Section~\ref{sec:dataModel}).
\item A frequency-based scoring methodology for searching personal digital traces. Our scoring, named {\em w5h-f} is based on our multidimensional data model and leverages entities interactions within and across dimensions in the data sets. (Section~\ref{sec:search})
\item An implementation of our techniques, from data extraction, to entity recognition, classification and index structures, that will be used as the basis of our experimental evaluation. (Section~\ref{sec:explSettings})
\item A thorough qualitative evaluation of our proposed {\em w5h} scoring and search techniques, as well as comparison with two popular existing search tools, Solr~\cite{ApacheSolr} and Spotlight~\cite{Spotlight}, and techniques, TFIDF~\cite{Salton:88} and BM25~\cite{Robertson96somesimple}, on real data using both manually designed and synthetically generated search queries. Our results show that our scoring model results in improved search accuracy. (Section~\ref{sec:experimentalEval}) 
\end{itemize}

We discuss related work in Section~\ref{sec:relatedWork} and conclude with future work directions in Section~\ref{sec:conclusion}.

\section{w5h Data Model}
\label{sec:dataModel}

We propose a data model that relies on the context in which personal data traces are created, produced and gathered to integrate heterogeneous traces into a unified data model that will support accurate searches. The proposed model, called \textit{w5h}, was derived from the following observations:

\begin{enumerate}
\item Personal digital traces are rich in contextual information, in the form of metadata, application data, or environment knowledge. 
\item Personal digital traces can be represented following a combination of dimensions that naturally summarize various aspects of the data collection: {\em who, when, where, what, why} and {\em how}.
\end{enumerate}

Our {\em w5h} model uses these six dimensions as the unifying features
of each personal digital trace object, regardless of its source. Using these
natural questions as the main facets of data representation will also
allow the combination of our data representation with a natural and
intuitive query model for searching information in digital traces.
Listed below
are some examples of dimensional data that can be extracted from a user's
personal digital traces:

\begin{itemize}
 \item \textbf{what}: messages, messages subjects, publications, description of events, description of users, list of interests of a user.   
\item \textbf{who}: user names, senders, recipients, event owners, lists of friends, authors.
 \item \textbf{where}: hometown, location, event venue, file/folder path, URL.
 \item \textbf{when}: birthday, file/message/event created-/modified-time, event start/end time. 
 \item \textbf{why}: sequences of data/events that are causally connected.
 \item \textbf{how}: application, device, environment.
\end{itemize}

\begin{figure}[ht!]
  \centering
   \includegraphics[scale=0.37]{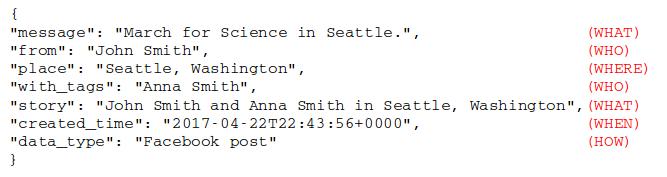}
  \caption[]{Simplified example of a user Facebook post classified according to the {\em w5h} model. }
  \label{fig:facebook_dim}
\end{figure}

Figure~\ref{fig:facebook_dim} presents a digital trace from a Facebook post with each piece of information identified as belonging to one of the six dimensions proposed (what, who, where, when, why and how). Even though multiple digital traces come from different sources and have their own data schema, they can be unified using the six dimensions proposed in our {\em w5h} model. For instance, two separate traces that have John Smith or/and Anna Smith under the same dimension {\em who} (for example a Facebook image tagging Anna Smith, or a tweet mentioning John Smith), can be linked by our unified model. Details on the implementation of the dimension classification and entity resolution are given in Section~\ref{sec:classification} and~\ref{sec:entityResolution}.  The {\em w5h} model is used both to unify heterogeneous digital trace data from different sources, and to link digital traces using the six proposed dimensions. 
 
The {\em why} dimension is not explored in this paper. This dimension can be derived by inference and could be used to connect different fragments of data. For instance, if a value could be inferred to the {\em why} dimension for the Facebook post in  Figure~\ref{fig:facebook_dim} it could be used to connect this data to a possible message thread. In~\cite{ExploreDBValia, odbase} we explored connections, in the form of plans, between events involving personal data traces; the plans, or tasks, connecting these events giving a contextual link as to {\em why} the corresponding digital traces were created/produced. 

In the next section we will explore indexing and searching techniques over sets of personal digital traces using the proposed {\em w5h} model .

\section{w5h Scoring Model}
\label{sec:search}
\label{sec:scoring}

We leverage the \emph{w5h} model presented in the previous section to provide rich and accurate search capabilities over personal digital traces. Unlike Web search, where the focus is often on discovering new
relevant information, search in personal data sets is typically
focused on retrieving relevant information that the user knows exists
in their data set.  In this scenario, standard search techniques are not
ideal as they do not leverage the additional knowledge
the user is likely to have about the target object, or the connections between objects pertaining to a given user.

 As pointed in~\cite{wagenaar86}, users tend to remember their
actions using the six natural questions; thus, using them to
guide search is a logical approach.  We now evaluate the potential benefits of the \emph{w5h} model for integrating and
searching personal data.  Specifically, we propose a search mechanism that supports queries containing conditions along each of the six interrogative dimensions. Our proposed search relies on a novel  frequency-based scoring methodology over the \emph{w5h} data model, called \emph{w5h-f}, that will be detailed in this section.

\subsection{Scoring Methodology}
\label{subsec:scoring}

To illustrate our query and scoring methodology let us consider the following search scenario: the user is interested in 
message(s) from John Smith or/and Anna Smith about the 2017 March for Science. We consider each digital trace to be a distinct object that can be returned as the result to a query.

\begin{definition}[Object in {\em w5h} Integrated Dataset]
\label{def:object}
An object $O$ in the data set is a structure that has fields corresponding to the 6 dimensions mentioned earlier. Each of these dimensions contains 0 or more items (corresponding to text, entities identified by entity resolution, times, locations, etc). The fields of an object $O$ are accessed using functions $O.get(``who"),$
$O.get(``what")$, etc.
\end{definition}

Formal queries have the same structure as objects in the unified data set. In the example above, the query has three filled dimensions: March for Science (\textbf{what}); John Smith, Anna Smith (\textbf{who}); 2017 (\textbf{when}).

Given objects \textit{Q} and \textit{O}, \textit{O} is considered as an answer to object \textit{Q} treated as a query if it contains at least one of the dimensions specified in \textit{Q}.  In looking for (partially) matching objects to a given query, each dimension will be searched separately, and the results will be combined according to a scoring function, generating a rank-ordered list of candidates. The choice of scoring function can be application dependent. We propose our frequency-based scoring function, {\em w5h-f}, below.

\subsection{w5h-f Scoring}
\label{subsec:w5hfscoring}

Because personal digital traces are byproducts of users' actions and events, they are not independent objects. Our intuition is that the correlation between traces (objects) can be leveraged to improve the accuracy of search results. For example, if the March for Science query from Section~\ref{subsec:scoring} returns several potential matches, one from Alice Jones, and one from Bob White, we may want to score the one from Alice higher if she communicates more frequently as a group with the user, Anna Smith, and John Smith, than Bob White.

 Our {\em w5h-f} scoring scheme uses the correlation between users (or entities) and how they interact 
over time to rank an object. Because we are focusing on personal digital traces, all the data articulates around a user.  By analyzing the data collected by our Extraction Tool~\cite{DataExtraction} (Section~\ref{sec:extractiontool}), we observed a strong correlation
 between the user (owner of the data) and multiple users ({\em who} groups), through times ({\em who, when}), location ({\em who, where}) and data sources ({\em who, how}). For instance, in one of the datasets, 94.9\% of the objects have more than 2 users ({\em who}), 95.7\% of objects have at least one date ({\em when}), 99.9\% of objects have content ({\em what}) and only 1.5\% of the objects have location ({\em where}). Our scoring exploits those interactions and correlations
 by  way of a frequency score. \footnote{Our model is focused around personal digital traces and as such we included this specific group of correlations in our scoring. Other application scenarios could also benefit from our {\em w5h}, with other group and pairwise correlations highlighted in a dedicated frequency-based scoring. For instance, traces from weather sensors could have strong pairwise {\em (where,when)}, or {\em (where, how)} correlations.}Frequencies can be computed for individual users or group 
 of users. They can be associated with multiple times, multiple data sources, and also with a set of locations.  For example, from a set of emails exchanged between a group of users, we can extract the frequency (number of interactions) with which those users communicated, and in which time period those interactions occurred. In short, frequency expresses the strength of relationships, based on users, time, location and data sources (\textit{who, when, where, how}).
 
  \begin{algorithm} [h!]
\caption{Frequency algorithm}
\label{FreqAlgorithm}
\begin{algorithmic}[1]
\Procedure{Compute\textendash Frequency(source)}{}
\State{\em\small /* object(source) retrieves all objects from a given source.}
\For{each O $\in$ object(source)}
	\State group $\gets$ O.get(`who')
	\State times $\gets$ O.get(`when')
	\State locations $\gets$ O.get(`where')
	\For {each time $\in$ times}
		\State $f[group][time] \gets f[group][time] + 1$  
		\For {each user $\in$ group}
			\State $f[user][time] \gets f[user][time]	 + 1$   
		\EndFor
	\EndFor
	\For {each user $\in$ group}
				\State $f[user] \gets f[user] + 1$  
	\EndFor
	\State $f[group] \gets f[group] + 1$ 	
	\For{each location $in$ locations}
		\State $f[location] \gets f[location] + 1$  
	\EndFor
\EndFor
\EndProcedure
\end{algorithmic}
\end{algorithm}

Algorithm~\ref{FreqAlgorithm} shows how frequencies are computed across multiple dimensions. Initially, a list of objects is retrieved for each data source. For each object, the algorithm extracts groups of users, times and locations. Then, the following frequencies are computed:
\begin{itemize}
\item Frequency of each individual user: number of objects that mention a user in the \textit{who} dimension.  
\item Frequency of a group of users: number of objects mentioning a group of users. If \{a,b,c\} is the group mentioned, frequencies of subgroups of \{a,b,c\}, e.g. \{a,b\} and \{b,c\}, are not counted.
\item Frequency of each individual user at specific times: number of objects that mention a user at matching times. Time is normalized, so variations are also considered. For instance, a query searching for June, will match objects with time June 2016 and June 2017.  
\item Frequency of a group of users at specific times: number of objects mentioning the group at a specific time. 
\item Frequency of a location: number of objects that mention a location. 
\end{itemize}
 
Besides computing the frequencies per source, we also compute the total frequency of a user, group of users, times and locations by combining the individual results obtained for each data source. For simplicity, in Algorithm~\ref{FreqAlgorithm}, every time a user or group of users has an interaction, the frequency is increased by one; however, in practice, the algorithm allows us to weigh differently distinct types of interactions. For example, likes or comments on a Facebook post could be weighed differently, giving more relevance to interactions coming from comments than likes. Different roles, e.g. From and To in an email, can also be weighed differently.

 \begin{definition}[Similarity Score]
  \label{def:frequency_score}
  Given a query \textit{Q}, an object \textit{O}, and the frequencies above, we define:

\begin{equation}
 \begin{split}
  \fscore(Q,O) &= f[g]  +  \sum_{u \epsilon who} f[u]  +  \sum_{u \epsilon who} f_{s}[u]   \\ 
                      &+ \sum_{\substack{u \epsilon who \\ dt \epsilon when}} f[u][dt]  + \sum_{\substack{u \epsilon who \\ dt \epsilon when}} f_s[u][dt]  \\
                      &+ \sum_{\substack{g \epsilon who \\dt \epsilon when}} f[g][dt] + \sum_{addr \epsilon where} f[addr]  \\ 
                      &+ score_{when}(dt, O) + score_{how}(s, O)  +  score_{what}(O) \nonumber
  \end{split}
\end{equation} 

  \noindent where $g$ is the group of users in the \emph{who} dimension of \textit{O}, $u$ is each user in $g$, $dt$ is
  each time in the \emph{when} dimension, $s$ is a data source, $addr$ is each location in the \emph{where}
  dimension, $f[g]$ is the frequency of a group of users in the same object, $f[u]$ is the total frequency of each  
  user across all data services, $f_{s}[u]$ is the frequency of each user in the data source $s$ of the object, $score_{when}(dt,O) = 1$ when the date $dt$ from query $Q$ matches object $O$; otherwise, $score_{when}(dt,O) = 0$, $f[u][dt]$ is the
  total frequency of the user $u$ in the time $dt$ across all data sources, $f_s[u][dt]$ is the frequency of the user $u$ 
  in the time $dt$ and data source $s$ of the object, $f[g][dt]$ is the total frequency of the group of user $g$ in the 
  time $dt$, $f[addr]$ is the frequency of each location $addr$, and $score_{how}(s, O)$ is the score of an object $O$ for a given source $s$:  $score_{how}(s,O) = 1$ when the service $s$ from query $Q$ matches object $O$; otherwise, $score_{how}(s,O) = 0$. Lastly, $score_{what}(O)$ is a text-based score for object $O$, using any chosen scoring function (e.g., TFIDF, BM25,...). 
\end{definition}

The equation in Definition~\ref{def:frequency_score} assumes that a query $Q$ has all 4 dimensions \emph{who, when, where} and \emph{how}; if a dimension does not exist in a query, the equation term corresponding to that dimension will be $0$.

Let us consider the query \textit{$Q_0$} (\textit{what:} March for Science; \textit{who:} John Smith, Anna Smith; \textit{when:} 2017), and the object \textit{$O_1$} illustrated in Figure~\ref{fig:facebook_dim} (Section~\ref{sec:dataModel}). According to the \textit{w5h-f} methodology, the object \textit{$O_1$} will have the following score: 

  \begin{eqnarray}
  \fscore(Q,O_1) &=& f[g=\text{John S., Anna S.}] \nonumber \\
  				   &+&  f[u=\text{John S.}] + f[u=\text{Anna S.}] \nonumber \\
  				   &+&  f_s[u=\text{John S.}] + f_s[u=\text{Anna S.}] \nonumber \\
  				   &+&  f[u=\text{John S.}][dt=\text{2017}] \nonumber \\
  				   &+& f[u=\text{Anna S.}][dt=\text{2017}] \nonumber \\
  				   &+&  f_s[u=\text{John S.}][dt=\text{2017}] \nonumber \\
  				   &+& f_s[u=\text{Anna S.}][dt=\text{2017}] \nonumber \\
  				   &+&  score_{when}(2017, O) \nonumber \\
  				   &+& f[g=\text{John S., Anna S.}][dt=\text{2017}] \nonumber\\
				   &+& score_{what}{``March for Science''} \nonumber
  \end{eqnarray} 
  \noindent where $s =$ Facebook

\section{Search Implementation}
\label{sec:explSettings}

We have presented a model to integrate personal digital traces into a unifying multi-dimensional data model in Section~\ref{sec:dataModel}. In Section ~\ref{sec:search}, we proposed a scoring methodology that leverages this data model to search heterogeneous data across all six dimensions while taking advantage of the inherent correlation between data objects in the scoring. We now discuss our search implementation in details.

\subsection{Data Retrieval}
\label{sec:extractiontool}

To create a data set of personal digital traces, we use the extraction tool proposed in~\cite{DataExtraction} to identify and retrieve data from current popular services and sources of digital traces. The data retrieved is stored in its original format to avoid mistakes that could lead to missing relevant data. All the data collected by the tool is stored in MongoDB, a NoSQL database that is already optimized for semi-structured data, with the data from each service stored in its own collection. We are constantly adding and revising sources of personal digital traces; the current implementation includes emails services (Gmail), social networks interactions (Facebook, LinkedIn, Twitter), location services (GPS, Foursquare), file management (Dropbox, Local Filesystem), browsing data (Firefox, Chrome), financial data (Mint, bank accounts), calendars (Google Calendar).

In the next section we will present how the raw data retrieved can be parsed and mapped into the {\em w5h} model proposed in Section~\ref{sec:dataModel}. 

\subsection{Classification}
\label{sec:classification}

Having defined the \textit{w5h} model (Section~\ref{sec:dataModel}), it is still necessary to find an effective mechanism to translate the heterogeneous set of personal data into the six dimensions. The dynamic nature of data sources, especially the rapid rate of change in the service APIs, and the fact that new sources can be added into the extraction tool, also pose a challenge.

Digital traces have their own structures but most are retrieved in a semi-structured data format (typically JSON through APIs), or are extracted along with some metadata. We implemented parsers to represent the raw data from each source in the \textit{w5h} model, thus unifying the data downloaded into a single data collection. The identification of data according to the six dimensions is done by analyzing the data available to be retrieved for each data source implemented and then building a dictionary of words/labels for each \textit{w5h} dimension. Much of the classification is intuitive, for instance, the words \textit{From} and \textit{To} should be classified under the \textit{who} dimension, while words \textit{Subject} and \textit{Body} should be classified as \textit{what}. Text messages are classified as \textit{what}, even though some specific information derived from content could be classified differently (e.g., ``I went to the market today'' gives both \textit{when} (``today''), \textit{where} (``market'') and \textit{who} (``I'')). Note that the \textit{how} and \textit{why} dimensions are more ambiguous. For now, we consider \textit{how} as the type of information recorded, e.g., a Facebook comment. The \emph{why} dimension is not explored in this paper; it is derived from inference and can be used to connect events~\cite{ExploreDBValia,odbase}.

\begin{figure}[ht!]
  \centering
   \includegraphics[scale=0.45]{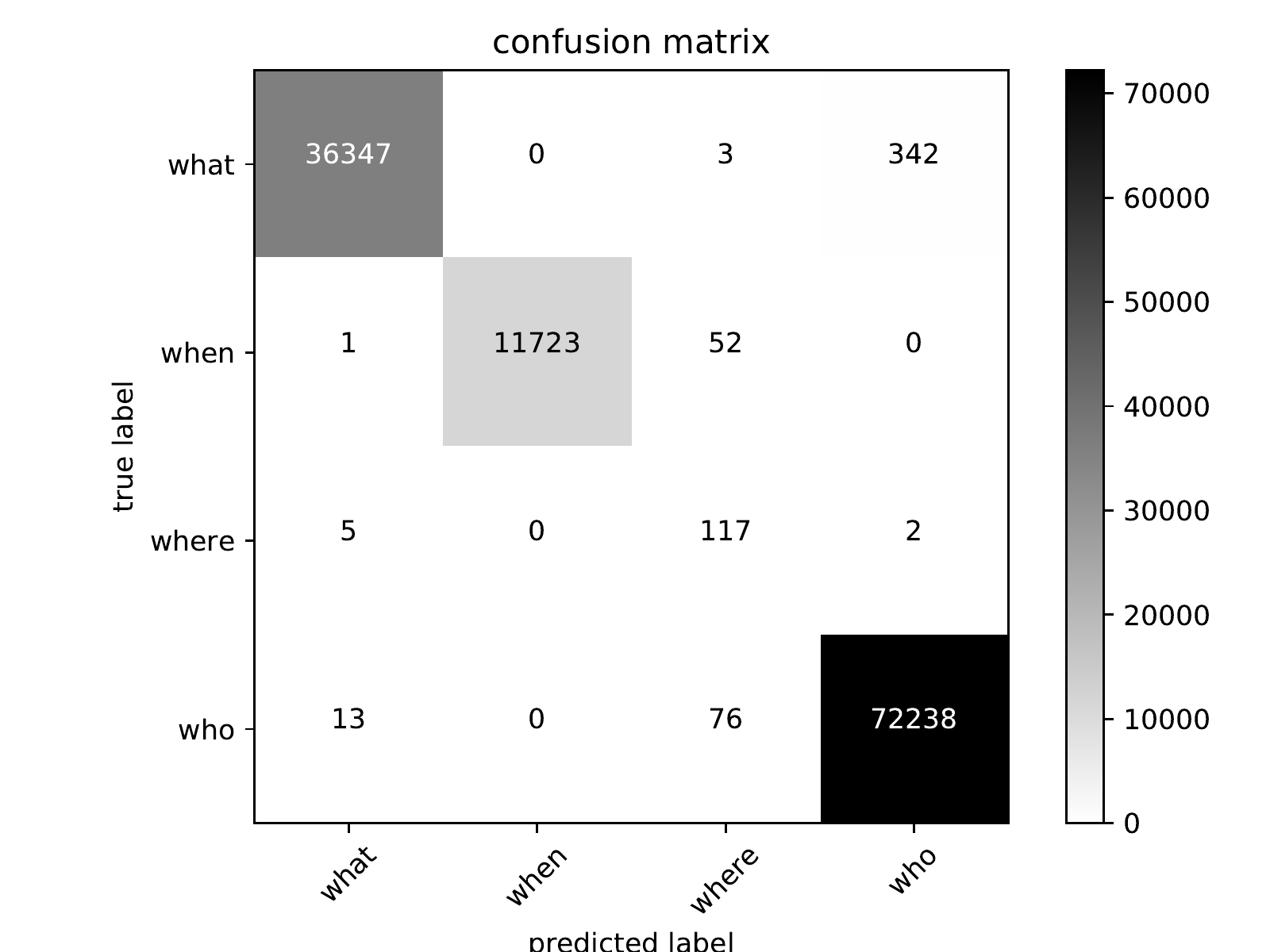}
  \caption[]{Confusion matrix with predictions for dataset \emph{User 1}. The model was trained using dataset \emph{User 2}. }
  \label{fig:confusionMatrix}
\end{figure}

We designed a machine learning multi-class classifier that automatically maps the raw data from each source into the \textit{w5h} dimensions. The input data to the w5h classifier is a set of sentences and \textit{w5h} labels. For instance, in Figure~\ref{fig:facebook_dim} (Section~\ref{sec:dataModel}) each line corresponds to a sentence/label pair. Each sentence is then transformed in embedding vectors by a Word2vec algorithm, and labels are reshaped into one-hot encoded binary matrices. Architectures were built combining LSTM (Long Short-Term Memory)~\cite{LSTM} and Dense layers. Dropout~\cite{dropout} was used in some architectures to reduce the complexity of the model with the goal to prevent overfitting. Parameters were evaluated using a 5-fold cross validation process to estimate the performance of models. We use categorical cross-entropy as the training criterion (loss function);  Adam optimization algorithm as the optimization algorithm for our models.   
The evaluation was conducted using the dataset \emph{User 2} described in~Table~\ref{table:data-statistics}. We achieve accuracy over 99.9\%. 
The confusion matrix in Figure~\ref{fig:confusionMatrix} shows the accuracy of the model for dataset \emph{User 1} (Table~\ref{table:data-statistics}), using the training data from \emph{User 2},  with the true labels represented in the y-axis and predicted labels in the x-axis. All correct predictions are located in the diagonal of the table. The results indicate that a machine learning classifier can accurately translate dynamic and heterogeneous set of personal data into the w5h model.

 Our implementation uses the classifier to translate raw data into the \textit{w5h} model and does not require user intervention.

\subsection{Entity Resolution}
\label{sec:entityResolution}

Our scoring technique (Section~\ref{sec:scoring}) relies on frequency scoring of the same entity across objects.
To make this possible, we need to identify separate instances of the same entity in data traces coming from the same sources, and across sources. For instance, the same person may appear in different services using variations of their names and email addresses{}. 

The impact of entity resolution on  search performance will be discussed in Section~\ref{sec:experimentalEval}.

\textbf{Entity Resolution for the \textit{who} dimension.} 
Almost 100\% of the personal data retrieved has information associated with the \textit{who} dimension. Our goal is to identify unique entities (person) that may be referred to differently (e.g. different email addresses{}). The first step to solve the ER problem for the \textit{who} dimension is to process the entire user data set, and extract all information classified under \textit{who}; for example, names and email addresses{}. We use the Stanford Entity Resolution Framework (SERF), a generic open-source, infrastructure for Entity Resolution
(ER)~\cite{SERF}, to identify entities. SERF uses the $swoosh$ algorithm~\cite{Benjelloun2009}, proved to be optimal in the number of record comparisons in worst-case scenarios. 
Using SERF person entities are identified and grouped in final entities that are stored in MongoDB in a separate collection.

\textbf{Entity Resolution for the \textit{where} dimension.} 
The same {\em where} location can be represented in multiple, ambiguous and error-prone ways. To disambiguate and match location data, we used Google Geocoding, Google Places API and SERF. We start by using Google Maps to disambiguate places that appear under different names and to augment the existing data. Besides dealing with multilingual places, Google Geocoding and Google Places API have the advantage of generating location-based data under the same format. For instance, Google Maps recognizes that Greece, Hellas, \textgreek{Ellada} and Grecia are the same location. However, there are a number of challenges to be faced. In most scenarios, given an ambiguous location (e.g. {\em Student Center}), the Google Maps API outputs a set of results instead of a unique address{}, making it difficult to identify which one of the listed addresses{} is the target place. To overcome this issue, we rank all addresses{} returned by a Google Maps search using a \textit{tf} (term frequency) function computed based on the user's data set. For example, consider a set of results returned by the API search; the set of addresses{} includes an address{} in France; if the user's data set does not have any data related to France, the address{} in France will be associated with a low \textit{tf}. Similarly, when Google Maps API does not return any result for a given search, we augment the location search by using information from other related digital traces. 
We then use SERF for deduplication and record linkage for all the locations that have the same geocoded address{} information or geographical coordinates (longitude, latitude).

\subsection{Retrieval}
\label{sec:index}

When a query is submitted, each dimension is individually matched against the user's data set using the above pre-computed indexes. Each separate search returns a list of objects  that partially match the query for a given dimension, which are then scored using the {\em w5h-f} scoring function (Section~\ref{subsec:w5hfscoring}). The current (unoptimized) implementation scores all matching objects and generates a ranked list of results. We are focusing our current efforts on validating the qualitative performance of our \emph{w5h-f} scoring model. We plan to investigate dedicated optimized \emph{w5h} index structures in the future.

\section{Experimental Evaluation}
\label{sec:experimentalEval}

We now evaluate the efficacy of the \emph{w5h-f} search approach by comparing its performance with two popular existing search tools, Solr~\cite{ApacheSolr} (using different scoring methodologies: TFIDF, BM25, and field-based BM25), and Spotlight~\cite{Spotlight}. In this section, we first describe our evaluation methodology. Then, we explore the accuracy of the search approach for a set of search scenarios manually designed to be representative of possible user queries. Finally, we explore the accuracy of the search approach using a much larger set of synthetically generated searches.

\subsection{Methodology}

\subsubsection{\bf Data Set.} 

\begin{table}[h]
\centering
\begin{tabular}{|l||r|r||r|r|}
\hline
                & \multicolumn{2}{c||}{User 1} & \multicolumn{2}{c|}{User 2} \\ \hline
\textbf{Data Source}     & \textbf{\#Objs}      & \textbf{Size}       & \textbf{\#Objs}      & \textbf{Size}           \\ \hline \hline
Facebook     	  & 1493  & 9Mb      & 2384    & 19Mb        \\ \hline
Gmail           	  & 1136  & 107Mb  & 10926  & 1Gb           \\ \hline
Dropbox       	  &    -      &      -       & 573      & 32Mb        \\ \hline
Foursquare   	  &     -     &       -      & 55        & 59Kb         \\ \hline
Twitter         	  &     -     &      -       & 2062    & 10Mb         \\ \hline
Google Calendar & 2        & 9Kb       & 209      & 389Kb        \\ \hline
Google+             & 1        & 1Kb       & 102      & 343Kb        \\ \hline
Google Contacts & 157    & 158Kb   & 427     & 430Kb         \\ \hline \hline
Total                  & 2789  & 116Mb   & 16738 & 1.4Gb          \\ \hline
\end{tabular}
\caption{Personal data sets for two users}
\label{table:data-statistics}
\vspace{-0.25in}
\end{table}

There is a dearth of
synthetic data sets and benchmarks to evaluate search over personal data. 
This challenge has only been exacerbated by the recent
explosion in the amount of personal digital traces, as well as the varied
services that create, collect, and store them.  Thus, we
perform our evaluation using a real data set collected by our extraction tool~\cite{DataExtraction} for two users. 

Table~\ref{table:data-statistics} shows two real user data sets along with the 
number and size of objects retrieved from different sources over different periods of time. These two data sets will be used to evaluate the \emph{w5h} scoring approach proposed in Section~\ref{sec:search}.

\subsubsection{\bf Evaluation Techniques.}\ \\

\textbf{Solr.} Solr~\cite{ApacheSolr} is a popular open source full-text search platform from the Apache Lucene project. For the experiments in this section, we integrate all data retrieved by the extraction tool, from each different data source, in a unified collection. This approach allows user to search for information across the entire set of retrieved digital traces, which is already a significant step forward from the current state. We consider three different scoring methods in conjunction with Solr: TFIDF, BM25, and field-based BM25 where the fields correspond to the parsing into the w5h model.

\textbf{Spotlight.} We also compare our search approach to Spotlight, the desktop search platform in Apple's OS X.  Spotlight allows users to search for files based on metadata~\cite{Spotlight}. This approach also works using the integrated raw (original) data. Each object in the evaluation data set is stored as an individual file in a machine running OS X Yosemite version 10.10.5. When possible, the following metadata is added to the files: MDAuthors (authors),  MDCreationDate (creation date), MDChangeDate (content change date), MDCreator (content creator),  MDFroms (path of a file). It is important to mention that Spotlight only ranks one item that it views as most relevant to a query.  All other matching items are returned without ranking, typically organized by type of documents (e.g., email, pdf, etc.).

\textbf{w5h-f} Our proposed approach relies on the six memory cues (what, who, when, where, why and how) to guide search. The \emph{w5h-f} approach uses the data parsed according to the \emph{w5h model}. The correlation between users/entities and how they interact over time through different services, including the frequency users communicate, is used to rank objects, as described in Section~\ref{sec:search}. \emph{w5h-f} uses entity resolution, as described in Section~\ref{sec:entityResolution},  to disambiguate/link entities from different sources (e.g. Facebook, Gmail, Twitter...) in the data set.

\begin{table*}[h!]
\begin{center}
  \small
  \begin{tabular}{|c|c|c|}
    \hline
    {\bf Search Approach} &  {\bf Query Description} & {\bf Rank}
    \\ \hline \hline
    \multicolumn{3}{|l|}{\vspace{-2.5mm}} \\
    \multicolumn{3}{|l|}{{\textit {Scenario 1 - search target: a Google+ post about SIGIR 2013 posted by Ashley in 2013}} } \\
    \hline
    Spotlight & MDContent: SIGIR, MDAuthors: Ashley, MDCreationDate: 2013  & 2 - 14 \\   
    Solr (TFIDF) & SIGIR, Ashley, 2013 & 11  \\ 
    Solr (BM25) & SIGIR, Ashley, 2013 & 12  \\
    Solr (Field-based BM25) & who:Ashley, what:SIGIR, when:2013 & 8 \\
    w5h-f & who:Ashley, what:SIGIR, when:2013 & 5  \\ 
    \hline \hline
    \multicolumn{3}{|l|}{\vspace{-2.5mm}} \\
   \multicolumn{3}{|l|}{{\textit {Scenario 2 - search target: a photo of
    a cat posted on Facebook by Katie in March 2012}} } \\ \hline
    Spotlight & MDContent:photo, MDContent:cat, MDAuthors:Katie, MDCreationDate:2012-03  & 2-2964 \\   
    Solr (TFIDF) & photo, cat, Katie, 2012-03 & 5468  \\ 
    Solr (BM25) & photo, cat, Katie, 2012-03 & 9106 \\
    Solr (Field-based BM25) & what:photo, what:cat, who:Katie, when:2012-03 & 65 \\
    w5h-f & what:photo, what:cat, who:Katie, when:2012-03 & 13  \\       
    \hline \hline          
    \multicolumn{3}{|l|}{\vspace{-2.5mm}} \\
    \multicolumn{3}{|l|}{{\textit {Scenario 3 - search target: a Facebook photo of Anna taken in Campos}} } \\ \hline
    Spotlight & MDContent:Photo, MDContent:Anna, MDContent:Campos & 2-3169  \\   
    Solr (TFIDF) & Photo, Anna, Campos &  17  \\ 
    Solr (BM25) & Photo, Anna, Campos & 43 \\
    Solr (Field-based BM25) & what: Photo, who: Anna, where: Campos & 1 \\
    w5h-f & what: Photo, who: Anna, where: Campos & 1 \\                         
    \hline                    
  \end{tabular}\par
\end{center}
\caption{Representative search scenarios targeting information stored in a
  user's personal data set.}
 \label{table:querySet1}
\vspace{-0.2in}
\end{table*}

\subsection{Case Studies}
\label{subsubsec:case1}

We begin our evaluation by studying three manually created search
scenarios designed to be representative of realistic user searches
targeting different personal digital traces from the
data set \emph{User 2} described in Table~\ref{table:data-statistics}.  For each scenario, we compose one query for each
of \emph{Spotlight}, \emph{Solr (TFIDF)}, \emph{Solr (BM25)}, \emph{Solr (Field-based BM25)} and \emph{w5h-f} using the
same information. Query conditions are derived from information in
the target objects, and all conditions are classified accurately along
the dimensions within \emph{Spotlight}, \emph{field-based Solr} and \emph{w5h-f}.

Table~\ref{table:querySet1} describes the search scenarios, the
corresponding queries, and the rank of the target object as returned
by each search method.  Note that the target objects are always found,
since the queries are accurate, and all three search tools currently
return all matching objects.  When Spotlight does not return the target item
as the 1st ranked result, we report the ranking as the range from 2 to
the total number of returned items.

The results show that \emph{w5h-f} achieves the best accuracy by always ranking the target object higher than or equal to {\em Spotlight} and {\em Solr}. The differences can be significant (e.g., scenarios 1, and 2), demonstrating that using memory cues to guide search can lead to improved search accuracy. We next discuss each of the search scenarios in more detail to show how differentiating between the dimensions, and using frequency information, helps to improve search accuracy.  

In scenario 1, the user is searching for a data item containing
information about the 2013 SIGIR Conference.  The information was
sent or posted by Ashley.  In this scenario,
identifying Ashley as {\em who} and 2013 as {\em when}
allows {\em w5h-f} to rank the target object higher than all instances of {\em
  Solr}. When compared with {\em Solr} field-based BM25, using the same parsed data as {\em w5h-f}, the fact that {\em w5h-f} scoring function takes into consideration the frequency that Ashley communicated with the user during the year of 2013 using Google+,  allows {\em w5h-f} to rank the target object higher than {\em Solr}. {\em Spotlight} was unable to leverage the same distinctions
as {\em w5h-f} since the target object was not ranked number
1.  Thus, {\em Spotlight} returned the target object as an unranked
item among 13 other items.

Scenario 2 targets a photo of a cat sent or taken by Katie in March 2012. In this case, the classification of photo and cat as {\em what} and Katie as {\em who} allows {\em w5h-f} and {\em Solr} field-based BM25 to rank the target object much higher than {\em Solr} BM25, {\em Solr} TFIDF and {\em Spotlight}. Entity resolution in the {\em who} dimension and the scoring function based on frequency help {\em w5h-f} to rank the target object in the top 20.

Scenario 3 looks for a picture of Anna taken at a place called Campos. The good performance achieved by the {\em w5h} and {\em Solr} field-based BM25 approach is explained by the fact that those approaches were able to classify Anna under the dimension \emph{who} and Campos under dimension \emph{where}. Since Campos is a very common family name in the user database, the keyword search approaches ended up returning lots of documents matching Campos as location and also as a name.

\begin{table*}[h!]
\begin{center}
\begin{tabular}{|l|l|l|l|l|l|l|}
\hline
{\bf Parameter} & {\bf Group 1} & {\bf Group 2} & {\bf Group 3} & {\bf Group 4} & {\bf Group 5}  \\ \hline
\hline
number of scenarios  & 250 & 250 & 250 & 250 & 250   \\ \hline
dimensions ($d$) & what & what, who & what, who, when & what, who, when, how & what, who, when, how  \\ \hline
number of values ($v$) & 1 & 1 & 1 & 1 & 2(who,what), 1(when,how)   \\ \hline
\end{tabular}
\end{center}
\caption{Parameters used to generate five groups of queries.}
\label{table:parameters_and_or}
\vspace{-0.2in}
\end{table*}

\begin{table}[h!]
\begin{center}
\begin{tabular}{| l | c | c | c |}
\hline
{\bf Methods} & {\bf MRR} & {\bf NDCG@10} & {\bf NDCG@20}  \\ \hline
\hline
Solr TF.IDF & 0.2920 & 0.3384 & 0.3673   \\ \hline
Solr BM25 & 0.4742 & 0.5192 & 0.5352   \\ \hline
Solr Field-based BM25 & 0.4979 & 0.5428 & 0.5619   \\ \hline
w5h-f (no entity) & 0.5632 & 0.5993 & 0.6136   \\ \hline
w5h-f & 0.6119 & 0.6414 & 0.6546   \\ \hline
\end{tabular}
\end{center}
\caption{MRR, NDCG@10, NDCG@20 for Group $2$ of queries.}
\label{table:entities}
\vspace{-0.1in}
\end{table}

\begin{table*}[h!]
\centering
\begin{tabular}{ccc}
\subfloat[Group 1]{
\begin{tabular}{| l | c | c | c |}
\hline
{\bf Methods} & {\bf MRR} & {\bf NDCG@10} & {\bf NDCG@20}  \\ \hline
\hline
Solr TF.IDF & 0.1959 & 0.2304 & 0.2513   \\ \hline
Solr BM25 & 0.2127 & 0.2481 & 0.2702   \\ \hline
Solr Field-based BM25 & 0.2383 & 0.2712 & 0.2996   \\ \hline
w5h-f & 0.2383 & 0.2712 & 0.2996   \\ \hline
\end{tabular}
} &
\subfloat[Group 3]{
\begin{tabular}{| l | c | c | c |}
\hline
{\bf Methods} & {\bf MRR} & {\bf NDCG@10} & {\bf NDCG@20}  \\ \hline
\hline
Solr TF.IDF & 0.3580 & 0.4036 & 0.4234   \\ \hline
Solr BM25 & 0.5267 & 0.5619 & 0.5777   \\ \hline
Solr Field-based BM25 & 0.6117 & 0.6582 & 0.6772   \\ \hline
w5h-f & 0.7072 & 0.7488 & 0.7628   \\ \hline
\end{tabular}
} \\
\subfloat[Group 4]{
\begin{tabular}{| l | c | c | c |}
\hline
{\bf Methods} & {\bf MRR} & {\bf NDCG@10} & {\bf NDCG@20}  \\ \hline
\hline
Solr TF.IDF & 0.3328 & 0.3925 & 0.4179   \\ \hline
Solr BM25 & 0.5357 & 0.5888 & 0.6036   \\ \hline
Solr Field-based BM25 & 0.6327 & 0.6765 & 0.6951   \\ \hline
w5h-f & 0.7539 & 0.7931 & 0.8013   \\ \hline
\end{tabular}
} &
\subfloat[Group 5]{
\begin{tabular}{| l | c | c | c |}
\hline
{\bf Methods} & {\bf MRR} & {\bf NDCG@10} & {\bf NDCG@20}  \\ \hline
\hline
Solr TF.IDF & 0.3772 & 0.4270 & 0.4569   \\ \hline
Solr BM25 & 0.5345 & 0.5924 & 0.6152   \\ \hline
Solr Field-based BM25 & 0.5769 & 0.6363 & 0.6510   \\ \hline
w5h-f & 0.6514 & 0.7014 & 0.7124   \\ \hline
\end{tabular}
} \\
\end{tabular}
\caption{MRR, NDCG@10, NDCG@20 for groups $1$,$3$,$4$,and $5$ (Group $2$ is in Table~\ref{table:entities}). Compared against w5h-f all the results are statistically significant (Wilcoxon signed-rank test).} 
\label{fig:groups}
\vspace{-0.25in}
\end{table*}

\subsection{Simulated Known-Item Queries}
\label{subsubsec:randomqueries}

We now study a larger set of automatically generated known-item queries: search of personal data is usually focused on retrieving information that users know exists in their own data set. Considering the fact that personal data trace search is a known-item type of search, simulated queries can be automatically generated, using known-item query~\cite{Elsweiler07towardstask-based} generation techniques such as the ones presented in ~\cite{Azzopardi:2007} and~\cite{Kim:2009}, as detailed below. 

For this set of experiments, we built two query sets, one using data set \emph{User 1}, and one using data set \emph{User 2} (Table~\ref{table:data-statistics}). Both sets comprise 5 different groups of queries, each containing 1500 queries for 250 different scenarios. Each scenario is automatically created by randomly choosing a target object from one of the evaluation data set. We then choose $d$ dimensions, from which we randomly select $v$ random values. We adapted the queries to each of our evaluation methods. 
Table~\ref{table:parameters_and_or} shows the parameters ($d, r, v$) for the 5 query groups. We performed our experiments on both \emph{User 1}, and \emph{User 2} data sets and observed similar behaviors. For space reasons, we only report here on the results over the \emph{User 2} data set.

Our evaluation resulted in the following observations on the impact of the multidimensional \emph{w5h} data model, choice of text search function, entity resolution, and frequency scoring on the accuracy of the search results.

\noindent \textbf{Including pertinent contextual information when searching personal data can significantly improve accuracy.} 
Tables~\ref{fig:groups} and~\ref{table:entities} show the MRR (Mean Reciprocal Rank), NDCG@10 (Normalized Discounted Cumulative Gain through position 10) and NDCG@20 (through position 20) of each approach, \emph{Solr} TFIDF, \emph{Solr} BM25, \emph{Solr} field-based BM25, and \textit{w5h-f}, for Group $1-5$ of queries. If the target object has the same ranking as other matching objects, we report the median value of the range. Observe that all search implementations that use the data parsed according to the \emph{w5h} model, \emph{Solr} field-based BM25, and \textit{w5h-f}, outperform the keyword-based approaches, \emph{Solr} TFIDF and \emph{Solr} BM25. These results show how valuable it is to use context (\emph{w5h-f} and \emph{Solr} field-based BM25) to find matching documents. 

\noindent \textbf{The use of a more elaborated approach to search text data can positively impact the final results obtained by the \textit{w5h} approaches.}
As previously mentioned, the \textit{what} dimension in the \textit{w5h} model is composed basically by content information comprising most of the text. \textit{w5h-f} uses \emph{Solr} field-based BM25 to score the \textit{what} dimension. The impact of the text search using \emph{Solr} field-based BM25 versus \emph{Solr} TFIDF and \emph{Solr} BM25, can be seen in Table~\ref{fig:groups} (a), which presents MRR, NDCG@10 and NDCG@20 for Group $1$ of queries (queries have only the \textit{what} dimension). We can observe that \emph{Solr} field-based BM25 and \textit{w5h-f} use a more efficient approach to search and score text data than \emph{Solr} TFIDF and \emph{Solr} BM25. Note that since Group $1$ has only one textual dimension in the query, the \emph{w5h-f} is equivalent to the underlying text-based scoring approach for the \textit{what} dimension; field-based BM25 in our implementation. The results  show that the adoption of a field-based text search for the \textit{what} dimension leads to better results.

\noindent \textbf{Being able to disambiguate/link people from different sources of data can significantly improve the accuracy of search.}
To analyze the importance of the entity resolution phase presented in Section~\ref{sec:entityResolution}, we created a group of queries (Group $2$) composed by values from the \textit{who} and \textit{what} dimensions. The results, for the data set \emph{User 2}, are illustrated in Table~\ref{table:entities}, with \emph{w5h-f} approach being superior when using entity resolution, compared with an implementation of \emph{w5h-f} that does not use entity resolution.

\noindent \textbf{Including frequency information as part of the scoring results in significant improvements.}
Tables~\ref{fig:groups} and~\ref{table:entities}  show that \emph{w5h-f}, which uses our proposed frequency scoring (Section~\ref{sec:search}), consistently outperforms \emph{Solr} field-based BM25, which also relies on the {\em w5h} model (Section~\ref{sec:dataModel}) but does not consider frequency. This shows that taking into consideration the correlation between dimensions while scoring an object improves the search accuracy. 

Our evaluation shows that using a tailored frequency-based multidimensional scoring approaches yields significant improvements in search accuracy over personal digital traces where the desired search outcome is a specific known object.

\section{Related Work}
\label{sec:relatedWork}

The case for a unified data model for personal information was made in~\cite{haystack,xu03towards}. deskWeb~\cite{deskweb} looks at the social network graph to expand the searched data set to include information available in the social network. Stuff I've Seen~\cite{stuff} indexes all of the information the user has seen, regardless of its location or provenance, and uses the corresponding metadata to improve search results. Seetrieve~\cite{Soules2008} extends on this idea by only considering the parts of documents that were visible to the user to infer task-based ({\em ``why''}) context to the file for later retrieval. Most notably, Personal Dataspaces~\cite{dataspace,iDM,imemex} propose semantic integration of data sources to provide meaningful semantic associations  that can be used to navigate and query user data (implicit context). Connections~\cite{soules2005} uses system activity to make similar connections between files; ~\cite{Soules2007} extends this approach to consider causality, using data flow, as contextual information. Our work is related to the wider field of Personal Information Management~\cite{PIMbook}, in particular, search behavior over personal digital traces is likely to mimic that of searching data over personal devices.  Unlike traditional information seeking, which focuses on discovering new information, the goal of search in Personal Information systems is to find information that has been created, received, or seen by the user. 

Bell has pioneered the field of life-logging with the project MyLifeBits~\cite{mylifebits,bell-total-recall} for which he has digitally captured all aspects of his life. While MyLifeBits started as an experiment, there is no denying that we are moving towards a world where all of our steps, actions, words and interactions will be recorded by personal devices (e.g., Google Glasses, cell phones GPS systems, FitBit and other Quantified Self sensors,...), or by public systems (e.g., traffic cameras, surveillance systems,...), and will generate a myriad of digital traces. {\em digi.me}~\cite{digime} is a commercial tool that aims at extending Bell's vision to everyday users. The motivations behind {\em digi.me} are very close to ours; however {\em digi.me} currently only offers a keyword- or navigation-based access to the data; search results can be filtered by service, data type or/and date.


Other file system related projects have tried to enhance the quality of search within the file system by leveraging the context in which information is accessed to find related information~\cite{chen09search, gyllstrom07confluence} or by altering the model of the file system to a more object-oriented database system~\cite{mic94file}.  YouPivot~\cite{YouPivot} indexes all user activities based on time and uses the time-based context to guide searches. Social context (users' friends and communities) is leveraged in~\cite{social-context} for information discovery; similarly~\cite{Derczynski2013} uses temporal and location context to aid discovery in social media data. Our work integrates all these sources of contextual information and provides a unified complete model of context-aware personal data.

Contextual information has been considered in various computer science applications. Context-aware applications dynamically adapt to changes in the environment in which they are running: location, time, user profile, history. Bolchini et al.\ provide a thorough survey of context-aware models in~\cite{ContextModelsSurvey}. Truong and Dustdar survey context-aware Web-Service systems in~\cite{Truong09asurvey}. Context-awareness has become increasingly popular with the wide adoption of mobile devices. While the types of context these systems consider overlap with ours, the overall approach is different from ours, for instance a contextually-aware Information Retrieval system will use the current context (e.g., user location and time of day) to adjust search results~\cite{Shen2005}. In contrast, we consider context as information that can be queried and used to guide the search.

\section{Conclusions and Future Work}
\label{sec:conclusion}

We proposed and implemented a multidimensional data model based on the six natural questions: \emph{what, when, where, who, why and how} to represent and unify heterogeneous personal digital traces. Based on this proposed model we designed a frequency-based scoring strategy for search queries that takes into account interactions between entities across objects to assist in the ranking of query results. Experiments over personal data sets composed by data from a variety of data sources showed that our approach significantly improved search accuracy when compared with traditional search methods. In the future, we plan on investigating several extensions to our work on searching personal data traces:
\begin{itemize}
\item Include topic modeling approaches over the \emph{what} dimension to be able to correlate objects based on their contents.
\item Optimize indexes and search algorithms to improve search efficiency.
\item Add query relaxation rules to allow for inaccuracies in the queries and approximate query matching.
\item Design an aggregate query model where {\em groups of objects (traces)} can be returned together as a query answer (e.g., all the social media messages and pictures relating to a party). For this we plan to integrate our work on the {\em why} dimension, which connects digital traces together~\cite{ExploreDBValia, odbase} into our scoring framework.
\end{itemize}


\bibliographystyle{abbrv}
\newpage
\bibliography{neemi}

\end{document}